# Antenna Model for Safe Human Exposure in Future 6G Smartphones: A Network Perspective

Haneet Kour, *Student Member, IEEE*, Rakesh Kumar Jha, Senior *Member, IEEE, Sanjeev Jain, Member IEEE*

***Abstract*— In this article we present the biological effect of antenna topology on a user's body. At different values of exposed frequency, the absorbent nature varies in human body. One of the major factors to be taken into consideration for designing 5G/6G mobile antenna is the biological effect and Electromagnetic Field Exposure (EMF). This is because the radio waves absorbed in human tissue leads to temperature elevation and has associated health hazards. We present an 8 port Multiple Input Multiple Output antenna system for use in 6G and beyond smartphone application devices. The proposed antenna is PIFA (Planar Inverted F Antenna) with dimensions 17.85 × $5mm^2$ and has dual band function at Sub 6GHz and 28/54 GHz band. The mean effective gain variation is < 3dB and the efficiency ranges from 65-75% for the proposed antenna. There are three case scenarios present for the proposed antenna system to reduce the EM radiation exposure. The execution of the proposed antenna system is studied in terms of EM metrics. The performance is validated with low peak SAR values obtained, low power density, exposure ratio and temperature elevation. There is EM radiation reduction while maintaining a desired Quality of service (QoS) and QoE. Battery life enhancement is also seen in a smartphone device with the proposed configuration.**

## I. INTRODUCTION

With the increasing ability of wireless communication industry to provide a wide variety of applications and services such as virtual reality, high-definition videos and so on, a breakthrough is seen in the design of 5G/6G networks. The goal of the future generation networks is to not only provide enhanced services but also ensure safety and reliability of operation. There are various technologies and architectures proposed in the literature to enhance the network ability to support large number of users with high bandwidth requiring applications. These technologies also optimize power and make the network energy efficient. Some of the most popular technologies include Device to Device communication, spectrum sharing (SS) that help in accommodating more number of users by utilizing the available spectrum efficiently and increasing the Quality of Service (QoS) and Quality of experience (QoE) of the end users [1]. As the 5G and 6G networks will be operating on very high frequencies such as millimeter-wave, it is imperative to study the impact of the Radio Frequency (RF) exposure of these waves on human body.

Haneet Kour and Sanjeev Jain are with Central University, Jammu, J&K, India. Rakesh Kumar Jha is with IIIDM Jabalpur, India.(Email:hani.kpds@gmail.com,jharakesh.45@gmail.com, dr_sanjeevjain@yahoo.com)

There is significant harmful impact of Electromagnetic (EM) Field on human tissues that is studied in the literature, which makes it crucial to study the biological effect and ensure that the safety standard for the future networks is maintained [2]. Various Electromagnetic radiation reduction techniques have been given to limit the harmful impact incurred from Electromagnetic waves. Some of them include Coordinated multipoint, RF shielding, base station switch off strategy, Massive MIMO (Multiple Input Multiple Output), change in antenna topology of the mobile phone [3-5]. The design of antenna to be used in future smartphones is also going to play a major role in limiting the harmful radiation exposure caused by the mobile device in hand, when used in talk mode (near the head) or in data mode (near the torso).
The mobile antenna configuration and hardware setup for one of the newest cellphones on the market today is shown in Fig. 1. It consists of the different mobile phone antennas and the major hardware components that form a part of the mobile phone with their locations specified as depicted in the figure.

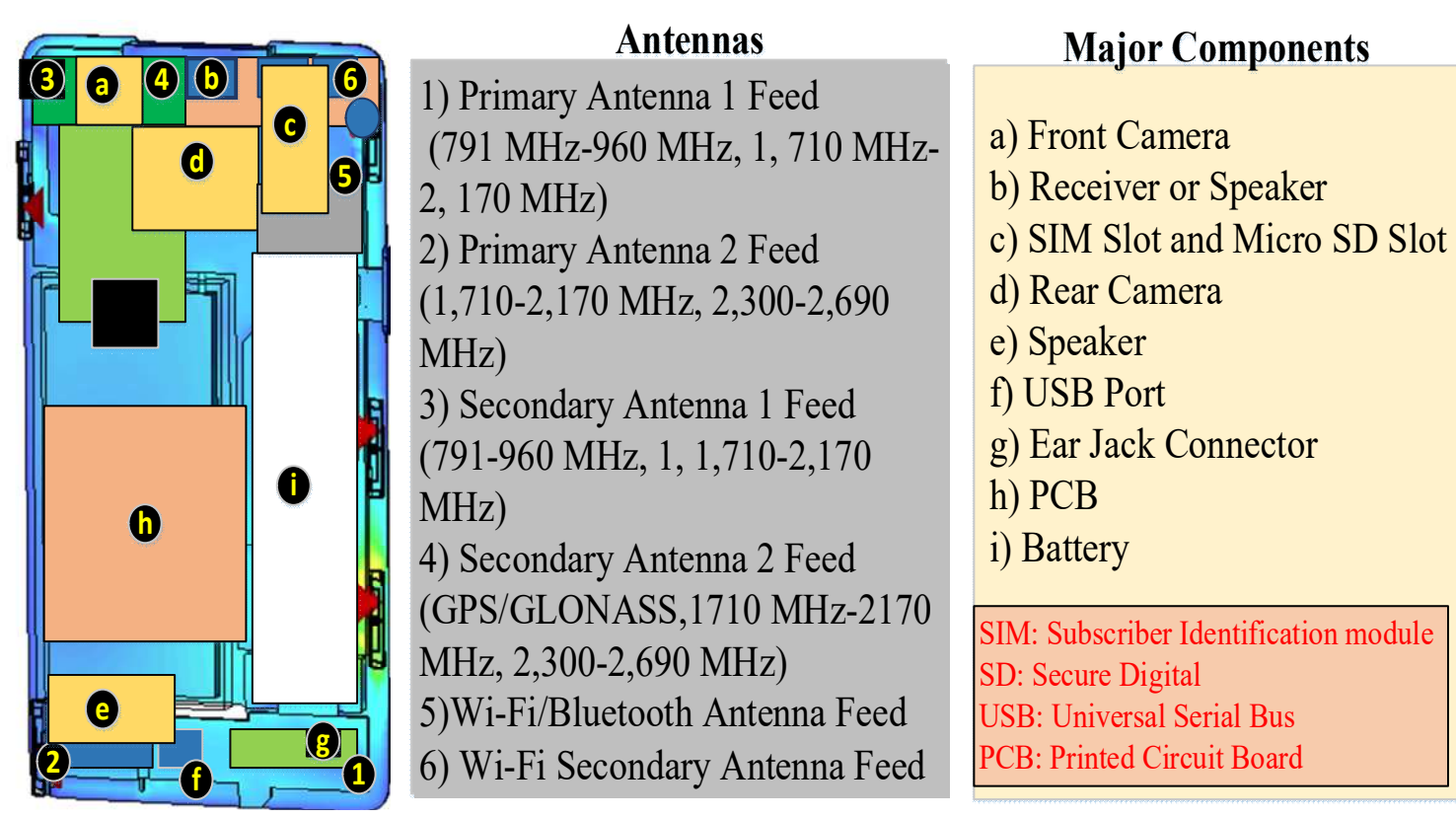

Fig. 1 A general mobile antenna and hardware configuration for Smartphones.

For the design of 6G mobile antennas there are some of the key parameters that have to be taken into consideration. One of these parameters include incorporation of antenna design that is most compatible with the existing technologies such as 2G/3G/4G/5G/GPS and Wi-Fi antennas [6]. Another parameter is integration of all the antennas to optimize the physical limitation of space. Various works in literature present the design of antenna elements for 5G smartphone applications for both the 5G New Radio (NR) bands i.e. Sub-6 GHz band and mmWave band. An antenna array is proposed for 5G MIMO application in the Sub 6-GHz band. It achieves high inter element isolation and radiation efficiency. Another dual polarized antenna for Sub 6-GHz 5G applications is proposed that achieves high average gain, antenna efficiency of 82.65 and provides radiation coverage for future 5G enabled handsets. For the mmWave band a four element MIMO antenna system is present in literature for future 5G systems which achieve a gain of 7.1 dBi. A nine element antenna array is proposed for mmWave 5G application achieving a very high efficiency of 93% and average gain of 20 dBi for 24-28.4 GHz band.

Studying the biological impact of EM field on human beings produced due to smartphone operation is another important parameter that pays a key role in design of 6G mobile antenna. There is requirement of antenna technologies that limit the incident power on the human body so that the local heating of tissues is minimized and there are no thermal hazards. There have been many advancements in the wireless industry with respect to the antenna and smartphone design parameters to incorporate operation at high frequencies such as sub-6-GHz and millimeter-wave (mmW) frequency bands. Incorporating MIMO and beamforming is among the most useful technologies for achieving high data rate with large spectral and energy efficiencies for 6G. The use of $4 \times 2$ MIMO in 4G smartphones has almost doubled the capacity of the network and this antenna array can be maximized for use in 6G with incorporating beamforming technology so that the attenuation losses or high path loss can be diminished. The frequency bands adopted in this work are the sub 6-Ghz frequency band and mmWave frequency bands i.e. Band 1: 0.54 GHz- 6GHz, Band 2: 24 GHz- 52 GHz. The smartphone size adopted is of the modern small size LTE smartphone with dimensions 70.5×145.7×7.8mm. In view of all the above parameters, there will a paradigm shift with respect to antenna technologies for mobile communication incorporating 5G and 6G generation.

### *A. Related Work*

Several works in literature discuss the importance of mitigating the EM radiation exposure caused due to smartphones. The authors in [7] discuss that the existing guidelines presented by FCC (Federal Communication Commission) are used to control the RF exposure caused by the current cellular networks. The SAR guidelines presented in these documents can be used to minimize the harmful radiation and local heating of tissues caused with the use of smartphones. Also, there are many governments that continue to rely on EMF guidelines are standards that were developed very early on in the journey of wireless communication. These guidelines need to be updated in accordance with the present status of ubiquitous mobile networks. It has been widely accepted that when antennas operate in close proximity to the body, there is a strong absorption of power that takes place in the human body tissues. The eyes and skin are most vulnerable to the heating and thermal radiation impact produced due to mmWave operation of devices. Investigations on antenna interactions with the human body at high frequencies such as 60 GHz have been studied in literature. The results have been demonstrated numerically as well as experimentally in the form of power density values and SAR rates for a certain input power [8].

To maintain the safety of operation of a mobile phone, the SAR levels should be well within the limits. It depends on various factors such as the type of antenna in operation, position of the antenna, phone style and also compactness of the antenna. The location of different antennas in the phone also plays a key role in the SAR produced by it such as the ground-free antennas must be placed at the bottom of the phone so that the SAR in head is less. In addition, the on-ground antenna such as PIFA (Planar Inverted F Antenna) must be located at the back of the phone for least SAR to be produced [9]. There are several proposals for the design of 5G mobile terminal antenna that can minimize the harmful impact even when operating at mmWave frequencies such as 28 GHz. The authors in [10] present the user effects when the mobile phone is operating in data use mode and talk mode. This is studied w.r.t different positions of the antenna i.e. at the top and bottom of the smartphone. It has been found that there is significant variation in the amount of power absorbed by the tissue with the change in location of antenna. 3D radiation patterns are presented for both the modes of operation.

In [11] the authors propose a $8 \times 8$ MIMO antenna that operates in the sub-6 GHz band for 5G smartphones. The position of the antennas is around the corners and sides of the phone. The performance of the fabricated antenna design is studied in terms of parameters such as Mean Effective Gain (MEG), channel capacity as well as SAR. The SAR patterns are obtained over 1g and 10g of tissue at frequencies of 3.55 GHz and 5GHz. The 3D radiation patterns are also obtained at the same frequencies. The harmful impact produced from a device can be reduced if the direction of the radiation beam is steered away from the head known as beam steering. The concept of beam switching antennas is found helpful in maximizing the gain of a mobile antenna and also enhances the coverage. It also helps in reducing the antenna profile i.e. the total number of Transmit-Receive (TR) components without compromising on spatial coverage. With reduction in the antenna profile along with beam switching and beam scanning, the overall exposure to the body tissues can be mitigated [12].

With the increase in frequency of operation, the size of the antenna also decreases. Developing an antenna that fulfills the property of providing high gain. High efficiency and compact in size is also a challenge for future mobile applications. In [13] an antenna element for 8 antenna MIMO system has been proposed that is compact in size and provides good efficiency for 3.5 GHz band. The size has been reduced by 30% with the help of vertical stub addition in the original antenna element. In [14] the authors propose dual antenna pair that also provides significant reduction in the size but achieves high antenna efficiency of about 74.7% and 57.8% for the dual antenna elements. In the current scenario most of the mobile phones have a full display screen. For such devices a multiple band antenna having very small non ground portion is widely popular. Keeping this in view, Nona-Band antenna is proposed in [15] for full display smartphones that accommodate the wireless frequencies till 4G LTE and also accommodate one band of 5G. It provides high efficiency and gain for the low band frequencies, middle band as well as high band frequencies.

The SAR reduction in MIMO systems can also be done by reducing the mutual coupling between the array of antennas [16]. The authors have suggested various decoupling mechanisms such as Defected Ground Structures (DGS), Dielectric Resonator Antenna (DRA) along with others to remove interference with focus on metamaterial and metasurfaces. Mutual coupling reduction is also studied in 34× 34 array antennas over Terahertz frequency range to improve gain and efficiency. An effective method for surface wave reduction is presented for 2 × 2 antenna array operating over different frequency bands. The mutual coupling between transmit and receive antenna scan also be reduced by using metamaterial photonic bandgap periodic structure. It operates on frequency range of 9.25-11 GHz achieving effective gain and reduces side lobes [17].

The motivation for this work is originally derived from our previous work published [18] in which we proposed half-duplex radio design for future mobile communication systems. The proposal was given to reduce the emitted EM radiation from a mobile phone so that the signals that are "always-on" in the network can be limited along with maintaining the desired QoS and QoE . The "Thermal Radiation" (TR) mode was proposed that limited the always-on uplink signals in the network so that the power radiated from a smartphone device is reduced. The detailed analysis of it is given in our publication [19] in which we present the simulation results of our proposed mode and the radiation reduction it achieves, in the form of EM metrics such as SAR (Specific Absorption Rate), power density and temperature elevation in the exposed body tissue.

### *B. Contributions and Organization:*

In this paper, an 8 port MIMO antenna configuration has been proposed for devices with 5G and beyond 5G applications. This proposal is present in context to biological safety of devices operating at high frequencies such as mmWave bands. The major offerings of this work are as follows:

- This article studies a major aspect required in the design of future mobile phone devices which is antenna design and its integration for devices with 5G and beyond 5G applications.
- We propose a MIMO antenna configuration to limit the Electromagnetic Field exposure produced from the active antennas in a mobile device with three case scenarios.
- The performance of the proposal is validated with simulation results in terms on EM radiation metrics such as SAR, power density, exposure ratio.
- The antenna configuration also helps in enhancing the battery lifetime of a device as we achieve the desired QoS and QoS while minimizing the power density radiated from the device in use.

The remainder of the article is organised as depicted ahead. The formulation of the problem and the system model are presented in Section II of the article. Model of the proposed antenna configuration is presented in Section III. Section IV provides the EM radiation metric calculation for the proposed methodology. In Section V, simulation results and analysis are presented. The final section of the article is Section VI in which the article concludes.

## II. PROBLEM FORMULATION AND SYSTEM MODEL

The problem formulation has been presented to analyze the impact of current MIMO systems for mobile communications with 4 antenna array and 8 antenna array system. Here we present the analysis for $4 \times 4$ MIMO and $8 \times 8$ MIMO in terms of current distribution, directivity, active impedance. In our analysis we have considered PIFA (Planar Inverted F Antenna) as it is the most widely used in modern smartphones due to the advantages associated i.e. high gain, high efficiency and appropriate radiation pattern. PIFA antenna also produces less Specific Absorption Rate (SAR) in comparison to other antennas and also has less interaction with hand-held environment. The analysis has been carried out at a frequency of 3.5 GHz i.e. in the 5G sub 6-GHz band. As we move towards higher frequency bands such as for 5G i.e. sub 6-GHz and mmWave, the wavelength decreases which in turn decreases the antenna size. PIFA has this associated advantage of having the smallest dimensions, which makes its integration compact in size and is very useful for slim structure of the smartphones that are being launched lately.

Fig. 2 depicts the MIMO system with 4 antenna array configuration It consists of the graphs depicting array layout as in Fig. 2 a), directivity in Fig. 2 b), current distribution in Fig. 2 c) and impedance graph in Fig. 2 d). The graph of directivity depicts the directionality of the radiation pattern of the array of antennas. It gives the maximum and minimum value of the directive gain. From the graph of directivity, it can be seen that the maximum value for a 4-antenna array system is 17.3 dbi. We desire to increase the directivity in a particular direction so that the antenna system has high gain and radiates less power to any users in its proximity. The current distribution for the antenna array system denotes the distribution of current from the feed point location i.e. the port of the antenna to the entire element. From the graph, we can see that the maximum value of current distribution is 800 mA/m at the feed point location and it decreases gradually along the element to a value of about 50 mA/m. Fig. 2 d) depicts the graph of antenna impedance.

It is calculated as the ratio of phasor voltage and phasor current at the port of the antenna. The graph here shows the active impedance of antenna element 1. Similar graph can be obtained for all the other antenna elements in this case i.e. 2,3 and 4.The graph for 8-antenna array system is depicted in Fig. 3. It also consists of array layout, directivity, current distribution and active impedance. Here the graph of directivity depicts a higher value of directional gain in comparison to that obtained for 4-antenna array system. As the number of antenna elements increase, the diversity increases and there is higher spatial multiplexing. This helps in achieving higher directivity for an antenna system with more number of antenna elements. The multiple antenna system can be used to increase the "diversity gain" as there is transmission of information in parallel streams. But the complexity of the system also increases with the increase in the number of antenna deployed.

Also, as we are studying the impact of safety of operation, the EM radiation exposure also increases with large number of antennas in operation. With respect to our proposed TR mode for mobile communication systems, we proposed that the number of uplink signals required for communication are much less than the downlink signals for communication. TR mode helps in limiting the uplink signal when it is in operation. TR mode has produced significant reduction in the amount of SAR, power density and temperature elevation as validated with the results obtained. We are studying the same impact with the antenna topology in smartphones. Fig. 2 and Fig. 3 depicts graphs for $4 \times 4$ and $8 \times 8$ antenna topology. As the number of uplink transmission of signals is much less in TR mode, the antennas responsible for uplink transfer of signals can be limited.

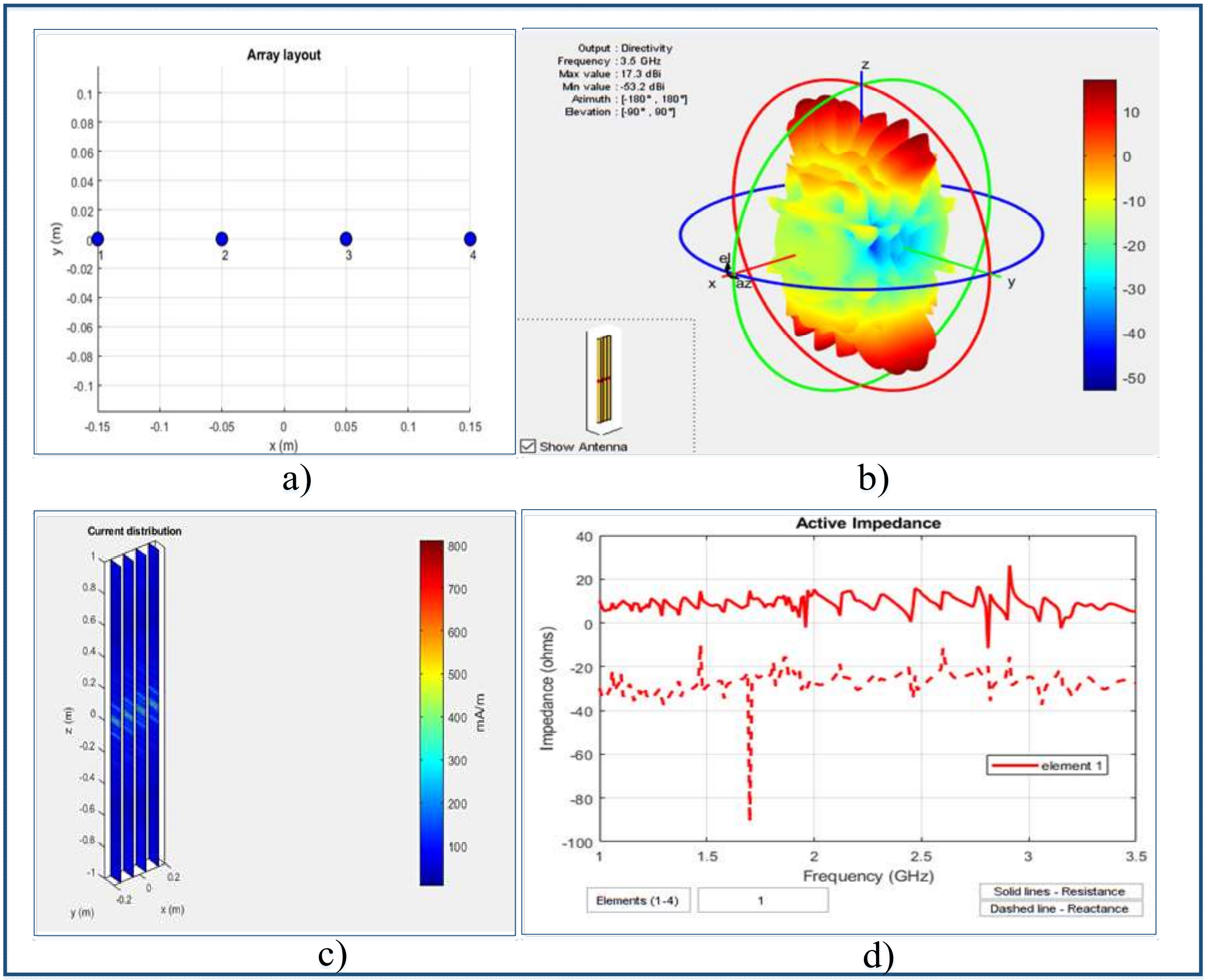

Fig. 2 MIMO system with 4-antenna array for smartphones.

To explain the concept of MIMO system model with equal number of transmit and receive antennas we can perform the case study as follows:

Case 1: Let us assume that the No. of Receive Antennas = No. of Transmit Antennas

Let us denote Receive Antennas by $r = 4$

No. of Transmit Antennas by $t = 4$

The channel equation can be denoted as follows:

$$y = Hx + n \qquad (1)$$

For a $4 \times 4$ system it can be written as

$$\begin{bmatrix} y_1 \\ y_2 \\ y_3 \\ y_4 \end{bmatrix} = \begin{bmatrix} h_{11} & h_{12} & h_{13} & h_{14} \\ h_{21} & h_{22} & h_{23} & h_{24} \\ h_{31} & h_{32} & h_{33} & h_{34} \\ h_{41} & h_{42} & h_{43} & h_{44} \end{bmatrix}_{r \times t} \begin{bmatrix} x_1 \\ x_2 \\ x_3 \\ x_4 \end{bmatrix} + \begin{bmatrix} n_1 \\ n_2 \\ n_3 \\ n_4 \end{bmatrix} \qquad (2)$$

The above equation can be combined and written in the following form

$$\bar{y} = H\bar{x} + \bar{n} \qquad (3)$$

Here $\bar{y}$ is the receive vector which is the $r$ dimensional vector. The channel matrix $H$ is the $r \times t$ dimensional vector. The vector $\bar{x}$ is the transmit vector which is $t$ dimensional and $\bar{n}$ is $r$ dimensional noise vector. The signal received at receiver 1 has the received component from all the transmitters that are active.

We require to decompose the MIMO channel $H$ so that receiver 1 receives signal from transmitter 1 only. We perform MIMO SVD i.e. Singular value decomposition. With this we perform decoupling of MIMO channels as follows:

$$\begin{bmatrix} \widetilde{y_1} \\ \widetilde{y_2} \\ \widetilde{y_3} \\ \widetilde{y_4} \end{bmatrix} = \begin{bmatrix} \sigma_1 & 0 & 0 & 0 \\ 0 & \sigma_2 & 0 & 0 \\ 0 & 0 & \sigma_3 & 0 \\ 0 & 0 & 0 & \sigma_4 \end{bmatrix}_{4 \times 4} \begin{bmatrix} \widetilde{x_1} \\ \widetilde{x_2} \\ \widetilde{x_3} \\ \widetilde{x_4} \end{bmatrix} + \begin{bmatrix} \widetilde{n_1} \\ \widetilde{n_2} \\ \widetilde{n_3} \\ \widetilde{n_4} \end{bmatrix} \qquad (4)$$

The decoded received signal vector can be written as follows for received signal at antenna 1

$$\widetilde{y_1} = \sigma_1 \widetilde{x_1} + \widetilde{n_1} \qquad (5)$$

The decoded signal received at receive antenna 2 can be written as

$$\widetilde{y_2} = \sigma_2 \widetilde{x_2} + \widetilde{n_2} \qquad (6)$$

The decoded signal received at receive antenna 3 can be written as

$$\widetilde{y_3} = \sigma_3 \widetilde{x_3} + \widetilde{n_3} \qquad (7)$$

The decoded signal received at receive antenna 4 can be written as

$$\widetilde{y_4} = \sigma_4 \widetilde{x_4} + \widetilde{n_4} \qquad (8)$$

The signal to Noise Ratio (SNR) can be computed for all the channels.

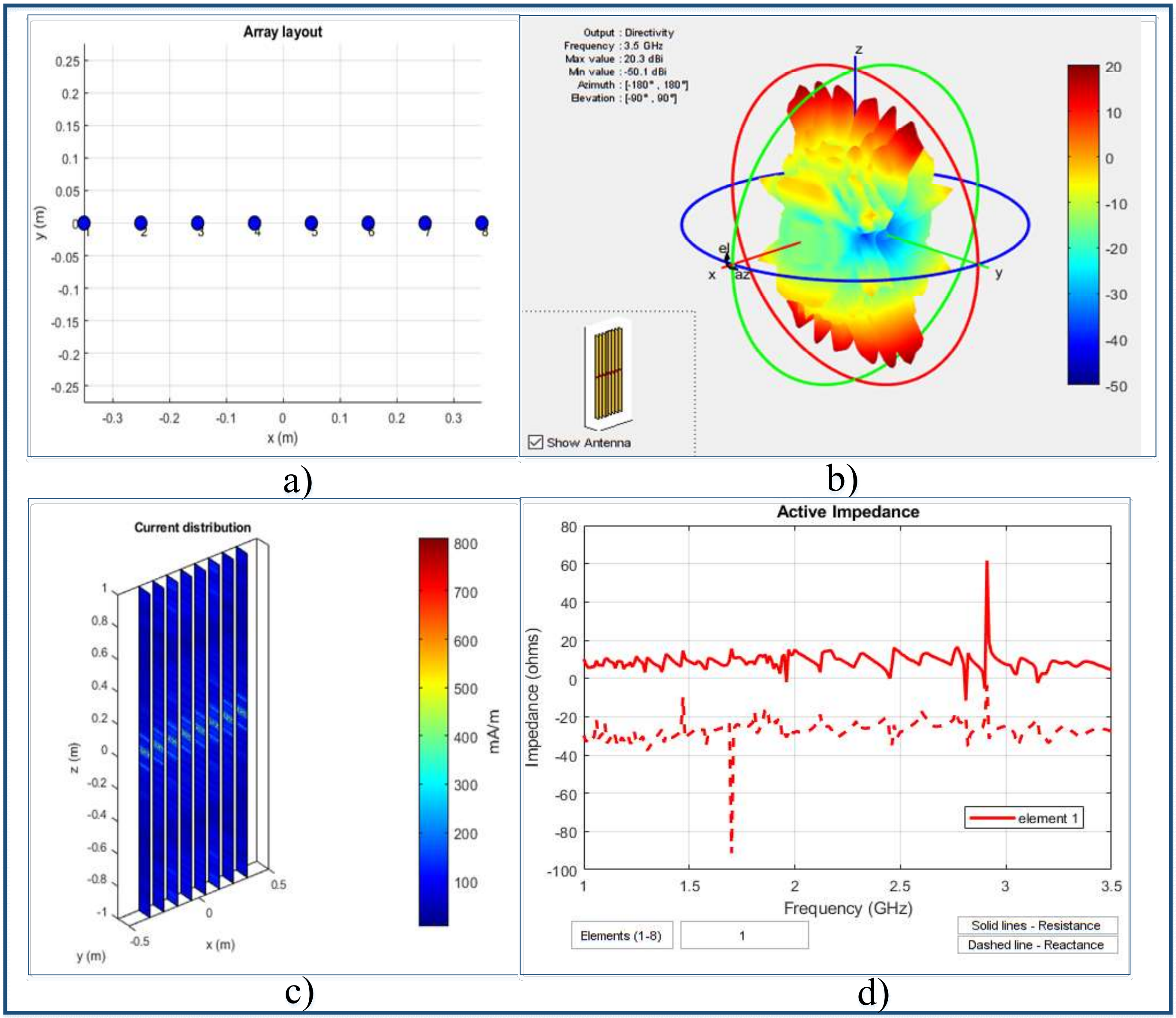

Fig. 3 MIMO system with 8-antenna array for smartphones.

The SNR for channel 1 is written as $\frac{\sigma_i^2 P_1}{\sigma_n^2}$, where $P_1$ is the transmitted symbol power for $x_1$. Since there are uplink as well as downlink signals from each antenna, we use power splitting ratio for uplink and downlink signals. The power splitting ratio is written as $\alpha_i \in [0,1]$. We denote the power consumed by uplink signals by $\alpha_i$ and by downlink signals with $(1-\alpha_i)$. The SNR obtained for Antenna 1 can be re-written as

$$SNR^{Ant1} = \frac{{\sigma_1}^2 {P_1}^{UL}}{\sigma_n^2} + \frac{{\sigma_1}^2 {P_1}^{DL}}{\sigma_n^2} \tag{9}$$

Since there is both uplink and downlink power consumed by antenna 1 and we have considered $4 \times 4$ antenna configuration, there are 4 UL signals and 4 DL signals. The equation (9) can be re-written as

$$SNR^{Ant1} = \frac{4\sigma_i^2}{\sigma_n^2}(\alpha_i P_1^{UL} + (1-\alpha_i)P_1^{DL}) \tag{10}$$

The ergodic capacity of a MIMO system can be calculated after computation of SNR as follows:

$$\overline{C_{erg}} = W \log_2 \det[1 + \frac{SNR}{N_t} H] \tag{11}$$

The ergodic capacity is used to compute the achieved target data rates for each of the antenna configuration. Here $W$ represents the bandwidth, $N_t$ is the number of transmit antennas and $H$ is the MIMO channel matrix. The calculated power for the individual antenna elements can also be used to calculate the value of directivity for every MIMO configuration.

The directivity for each of the antennas can be computed with the following expression

$$Directivity\ (D) = \frac{U_{given\ direction}}{U_{average}} \tag{12}$$

Here $U_{given\ direction}$ is the radiation intensity produced from antenna 1 in a given direction. $U_{average}$ is the radiation intensity average over all directions. The average radiation intensity is obtained as follows:

$$U_{average} = \frac{P_{radiated}}{4\pi} \tag{13}$$

We are considering the radiated power caused only by the uplink signals because they contribute most to the radiation exposure as they are directly incident on human body tissue. The uplink signals cause maximum heating of the device and lead to a high battery power consumption in a short duration of operation.

On substituting the above equation in the equation for directivity, it can be written as

$$Directivity\ (D) = U_{given\ direction} \times \frac{4\pi}{P_{radiated}} \quad (14)$$

The value of power radiated from an antenna is more when the number of transmit and receive antennas are large. In this case the total uplink power can be written as $4\alpha_i P_1^{UL}$. Substituting the uplink power in equation (13), it can be written as

$$Directivity_{(4\times4)} = U_{given\ direction} \times \frac{4\pi}{4\alpha_i P_1^{UL}} \quad (15)$$

The value of $U_{giv\quad direction}$ can be calculated at a particular value of elevation and azimuthal angle. The value of $U_{given\ direction}$ can be computed with the following expression

$$U_{given\ direction} = W \times r^2 \quad (16)$$

where $W$ is the power density that can be calculated from the electric field intensity obtained by a particular field pattern formed by an antenna. The final expression for directivity is as follows:

$$Directivity_{(4\times4)} = (W \times r^2) \times \frac{4\pi}{4\alpha_i P_1^{UL} + (1-\alpha_i) P_1^{DL}} \quad (17)$$

Case 2: Let us assume that the No. of Receive Antennas ≠ No. of Transmit Antennas

Let us denote Receive Antennas by $r = 4$

No. of Transmit Antennas by $t = 2$

The channel equation can be denoted as follows:

$$y = Hx + n \quad (18)$$

For a 4 × 2 system it can be written as

$$\begin{bmatrix} y_1 \\ y_2 \\ y_3 \\ y_4 \end{bmatrix} = \begin{bmatrix} \sigma_1 & 0 \\ 0 & \sigma_2 \\ 0 & 0 \\ 0 & 0 \end{bmatrix}_{4\times2} \begin{bmatrix} \widetilde{x_1} \\ \widetilde{x_2} \end{bmatrix} + \begin{bmatrix} n_1 \\ n_2 \\ n_3 \\ n_4 \end{bmatrix} \quad (19)$$

The above equation can be combined and written in the following form

$$\bar{y} = H\bar{x} + \bar{n} \quad (20)$$

Again $\bar{y}$ is the receive vector which is the $r$ dimensional vector. The channel matrix $H$ is the $r \times t$ dimensional vector. We perform decoupling of MIMO channels as follows:

$$\begin{bmatrix} \widetilde{y_1} \\ \widetilde{y_2} \\ \widetilde{y_3} \\ \widetilde{y_4} \end{bmatrix} = \begin{bmatrix} h_{11} & h_{12} \\ h_{21} & h_{22} \\ h_{31} & h_{32} \\ h_{41} & h_{42} \end{bmatrix}_{r\times t} \begin{bmatrix} \widetilde{x_1} \\ \widetilde{x_2} \end{bmatrix} + \begin{bmatrix} \widetilde{n_1} \\ \widetilde{n_2} \\ \widetilde{n_3} \\ \widetilde{n_4} \end{bmatrix} \quad (21)$$

The decoded received signal vector can be written as follows for received signal at antenna 1

$$\widetilde{y_1} = \sigma_1 \widetilde{x_1} + \widetilde{n_1} \quad (22)$$

The decoded signal received at receive antenna 2 can be written as

$$\widetilde{y_2} = \sigma_2 \widetilde{x_2} + \widetilde{n_2} \quad (23)$$

The decoded signal received at receive antenna 3 can be written as

$$\widetilde{y_3} = \widetilde{n_3} \quad (24)$$

Since it is a 4 × 2 antenna system, there are only two transmit data symbols and four receive vector channels.

The decoded signal received at receive antenna 4 can be written as

$$\widetilde{y_4} = \widetilde{n_4} \quad (25)$$

The signal to Noise Ratio (SNR) can be computed for all the channels.

The SNR for channel 1 is written as $\frac{\sigma_i^2 P_1}{\sigma_n^2}$, where $P_1$ is the transmitted symbol power for $x_1$. Similarly, the SNR for channel 2 can be formulated as $\frac{\sigma_2^2 P_2}{\sigma_n^2}$.

Since there is a reduction in the number of transmit channels in this case, the power factor for channel 3 and 4 does not exist as there is no information transmitted on these paths. This reduces the total power component radiated from the device with 4 × 2 antenna configuration.

The directivity in this case will be produced only from the two transmitting antennas. It can be computed as follows for the first transmit antenna:

$$D_{4\times2}(Ant1) = U_{given\ direction} \times \frac{4\pi}{P_1} \quad (26)$$

For the second transmit antenna, the directivity is calculated as follows:

$$D_{4\times2}(Ant2) = U_{given\ direction} \times \frac{4\pi}{P_2} \quad (27)$$

The directivity for the two transmit antennas in the case of 4 × 2 antenna system can be computed from equation (25) and (26). The total directivity from the two transmitting antennas can be written as

$$D_{4\times2}(Total) = U_{given\ direction} \times 4\pi \left(\frac{1}{P_1} + \frac{1}{P_2}\right) \quad (28)$$

The obtained values of power from the antennas in both the case scenarios can be used to calculate the values of SAR, power density and other EM metrics as explained in Section IV. In the following section we present our proposed antenna configuration for 5G and beyond 5G smartphones to limit the EM radiation impact.

## III. Proposed Antenna Configuration

We propose offloading the network traffic by reducing the number of transmit antennas in the smartphone when there is less requirement of all uplink signals to be active in the network. The proposal is presented for an 8-antenna array system in which there are 4 UL and 4 DL channels conventionally. As it has been observed that in the current communication scenario and in the future mobile systems that maximum communication occurs over the Wi-Fi network [20] for indoor communication. In this case the number of transmit antennas that are always-on can be reduced in the smartphone for some duration of time.

When the smartphone communicates over Wi-Fi for internet consumption there is no requirement of always-on signals in the network. Some of the antennas responsible for uplink transfer of information can be offloaded at this point as the uplink signals are mostly responsible for causing high power density and SAR in the mobile phone. When we reduce the number of transmit antennas that are active in the mobile phone, the total power radiated decreases and in turn the directivity increases as per the equation of directivity. The simulated configuration of antenna design for our proposed topology is depicted in Fig. 4 (a) and (b). We depict the current 8 Antenna system with always-on antennas i.e. 4 UL (Uplink) and 4 DL (Downlink) antennas and proposed antenna system with OFF antennas respectively.

The design is constructed using CST Microwave Studio having substrate of 0.8 mm thickness, dielectric constant of 2.8 and tangent delta of 0.002.

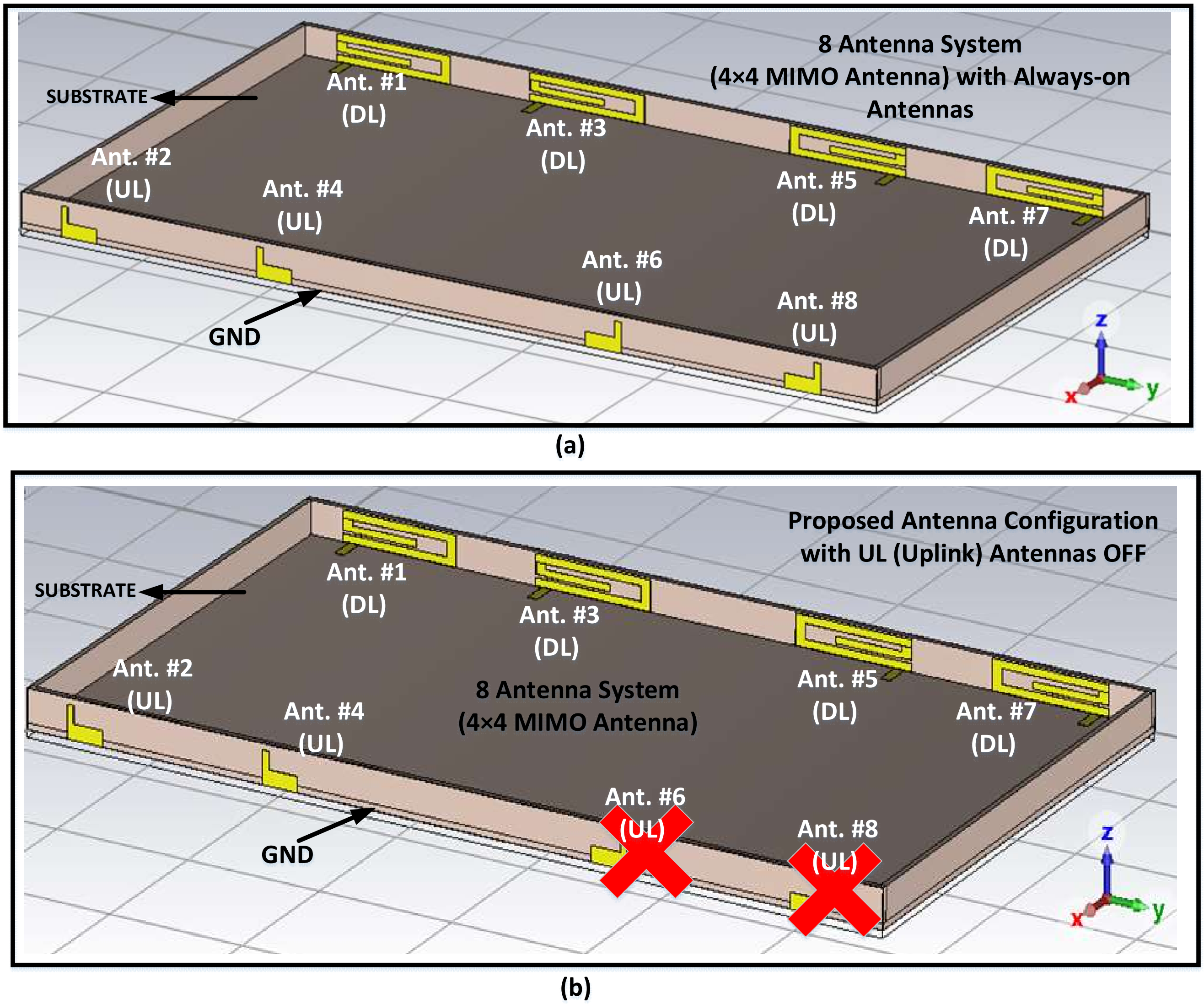

Fig. 4 Simulated Configuration of (a) current antenna design and (b) proposed Antenna design.

The substrate has ground plane which has antenna array with side frames having dimensions 3.9mm × 17mm. In the proposed design (b) the two UL antennas that are offloaded reduce the EM radiation emission without compromising on the QoS and QoE of a user in the event of a mobile device communicating over Wi-Fi. There can be different types of antenna design such as H-Shaped Eight -element dual-band MIMO Antenna, folded bowtie Antenna, Uni-planar MIMO Antenna as discussed in Section I. Further in our proposed antenna configuration, we present three case scenarios as depicted in Fig. 5 and explained as follows:

*Case 1: Offloading Two Transmit Antennas*

Currently the 8-antenna array consists of 4 antennas for downlink and 4 for uplink. In this case we propose offloading of two transmit antennas i.e. Uplink antennas when the mobile phone user is communicating over Wi-Fi for internet consumption. During this period the uplink transfer signals are required only to maintain a stable connection with the base station that is required in case of any call set up procedure to be established. When we offload the two transmit antennas, the number of active antennas now include the downlink 4 antennas and remaining two transmit antennas. All the antennas operating in the 5G frequency band are not active. Some are offloaded to limit the emitted radiations.

Now the active antennas in this case include the Wi-Fi antennas, 4 downlink antennas and 2 uplink antennas. With this topology there is achievement of required data rate by the user, stable connection maintained with the base station as well as reduction in the emitted radiations. The individual diagrammatic representation of case 1 is shown in Fig. 6 with conventional and proposed configuration. In the existing configuration we have alternate uplink and downlink signals. In the proposed scheme two transmitting antennas are depicted as offloaded or in inactive mode when the device is using internet over Wi-Fi. The reduction in active transmit antennas reduces the radiated emissions from the device. Uplink signals are the major contributors of thermal heating caused in the device as well as radiated emissions. It leads to high specific absorption rate and local heating of the exposed tissue. When there is internet consumption over Wi-Fi, the received SNR by the smartphone is more over the Wi-Fi in comparison to the SNR obtained over mobile data.

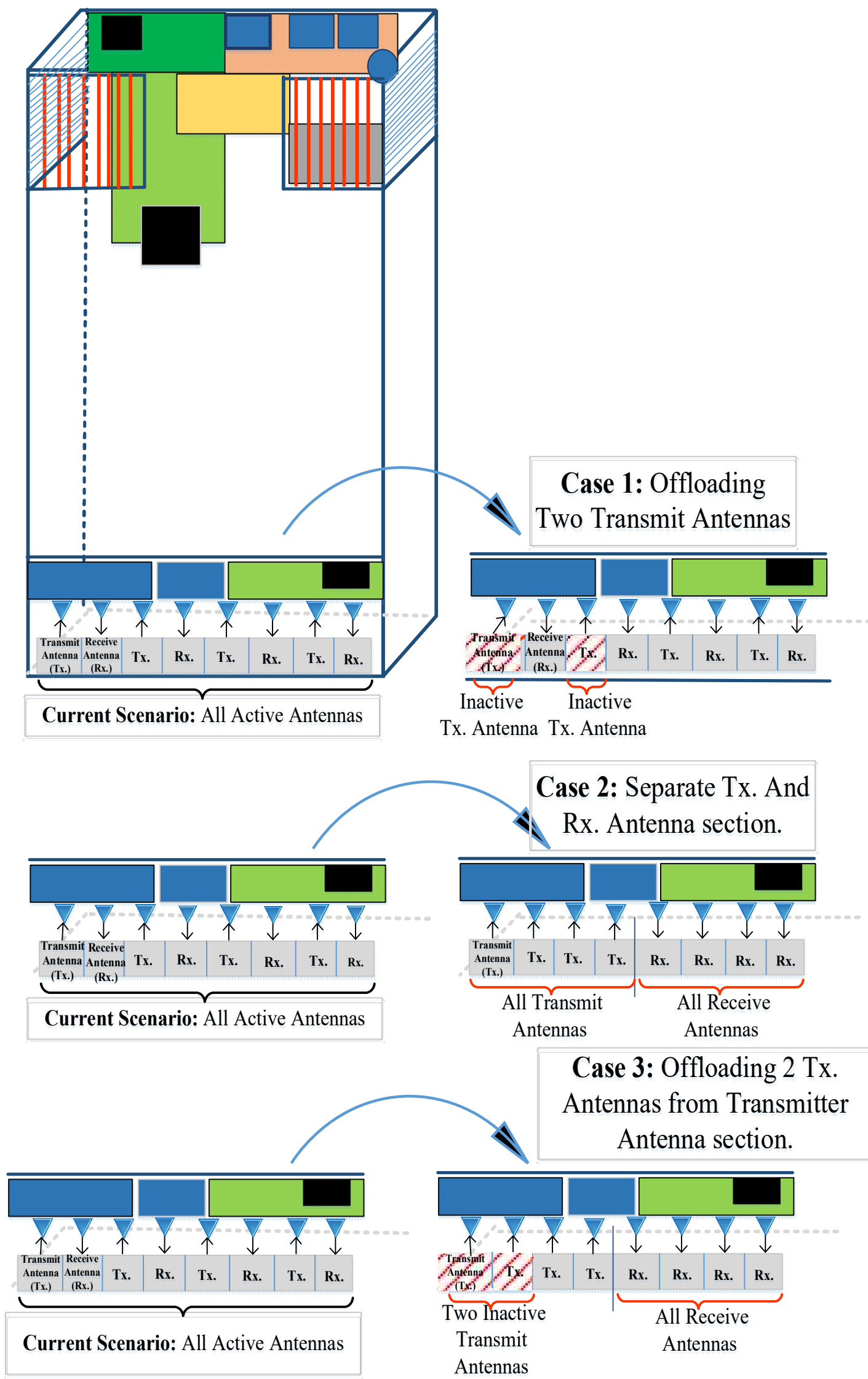

Fig. 5 Case Study analysis for proposed antenna configuration.

During this process there is maintenance of a stable connection with the base station at all times in case of any call set-up procedure that is required at any point. A flowchart depicting the algorithm for case 1 is presented in Fig.7.

It depicts the setup procedure for a mobile device consuming internet over Wi-Fi or mobile data. When the RRC (Radio Resource Control) setup is complete with the gNB a stable connection is established with the base station. After that it is checked if the internet is ON. If the device is communicating over Wi-Fi the flag status is changed to 1 and Wi-Fi antennas are active. If the device is communicating on mobile data, then the flag is inactive i.e. 0. For a device communicating over Wi-Fi the received SNR is checked and is found to be higher than that achieved over mobile data. This received higher SNR helps in achieving higher data rate. At this instance we propose offloading of 2 Transmit antennas. This helps in decreasing the number of active antennas that are ON for a particular duration which will further decrease the utilization of power by the device.

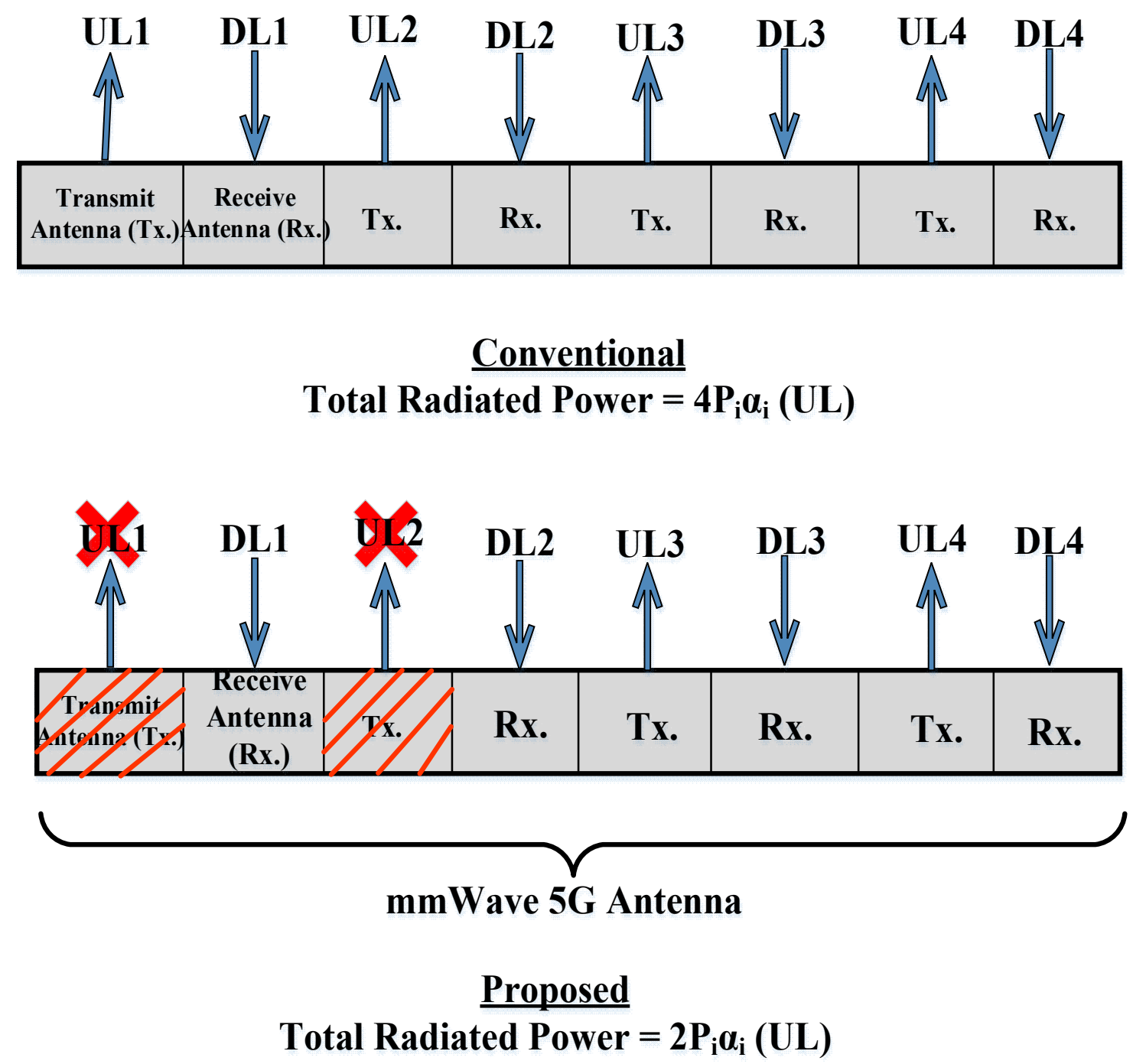

Fig. 6 Proposed Case scenario 1 analysis.

This decrease in power consumption is directly associated to the power density radiated from a device and the SAR produced in the directly exposed tissue of the body. Also, for the duration the uplink signals are switched off, the battery consumption by the mobile device also reduces significantly.

*B. Case 2: Separate Transmit and Receive Antenna Section*

In this case the transmit and receive antennas in an antenna layout have been divided based on uplink or downlink information transfer. The current antenna layout consists of alternate transmit and receive antennas placed alongside. This layout causes interference between UL-DL signals in the network. We propose dividing the layout with one section containing all the transmit antennas and the other section containing all the receive antennas. With this layout, there will be reduced interference between all the uplink and downlink signals. The interference will be caused at only one junction in the middle, where one transmit and one receive antennas are placed alongside. All the other antennas do not cause any interference in the transfer of information to and from the mobile device.

This is another method to limit the radiated power density from a smart phone device. The diagrammatic representation of the proposed case scenario 2 is presented in Fig. 8. The total interference caused in the conventional scenario is an arithmetic addition of interference between each transmitter and receiver unit. It can be formulated as follows:

$$I_c^{total} = I_{12} + I_{23} + I_{34} + I_{45} + I_{56} + I_{67} + I_{78} \quad (29)$$

In the proposed case scenario, the total interreference produced is as follows:

$$I_c^{proposed} = I_c^{total} - (I_{12} - I_{23} - I_{34} - I_{56} - I_{67} - I_{78}) + I_{45} = I_{45} \quad (30)$$

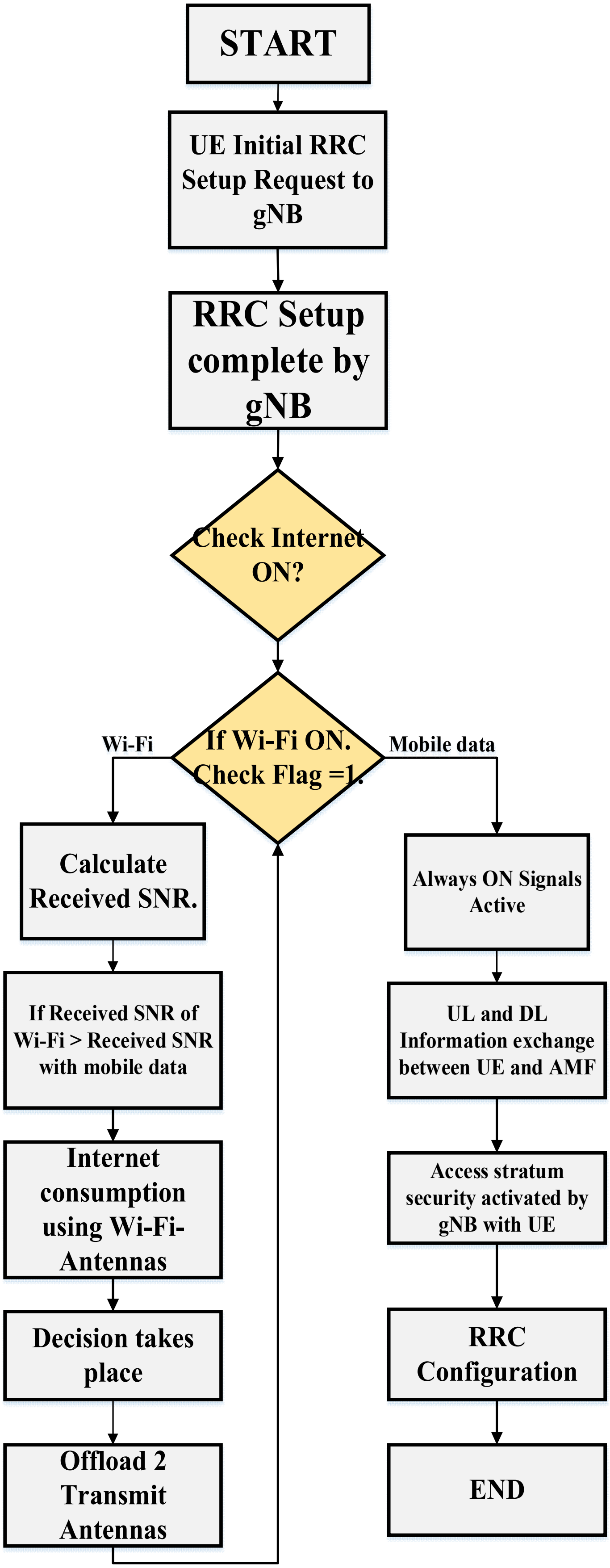

Fig. 7 Algorithm for proposed antenna configuration for case 1.

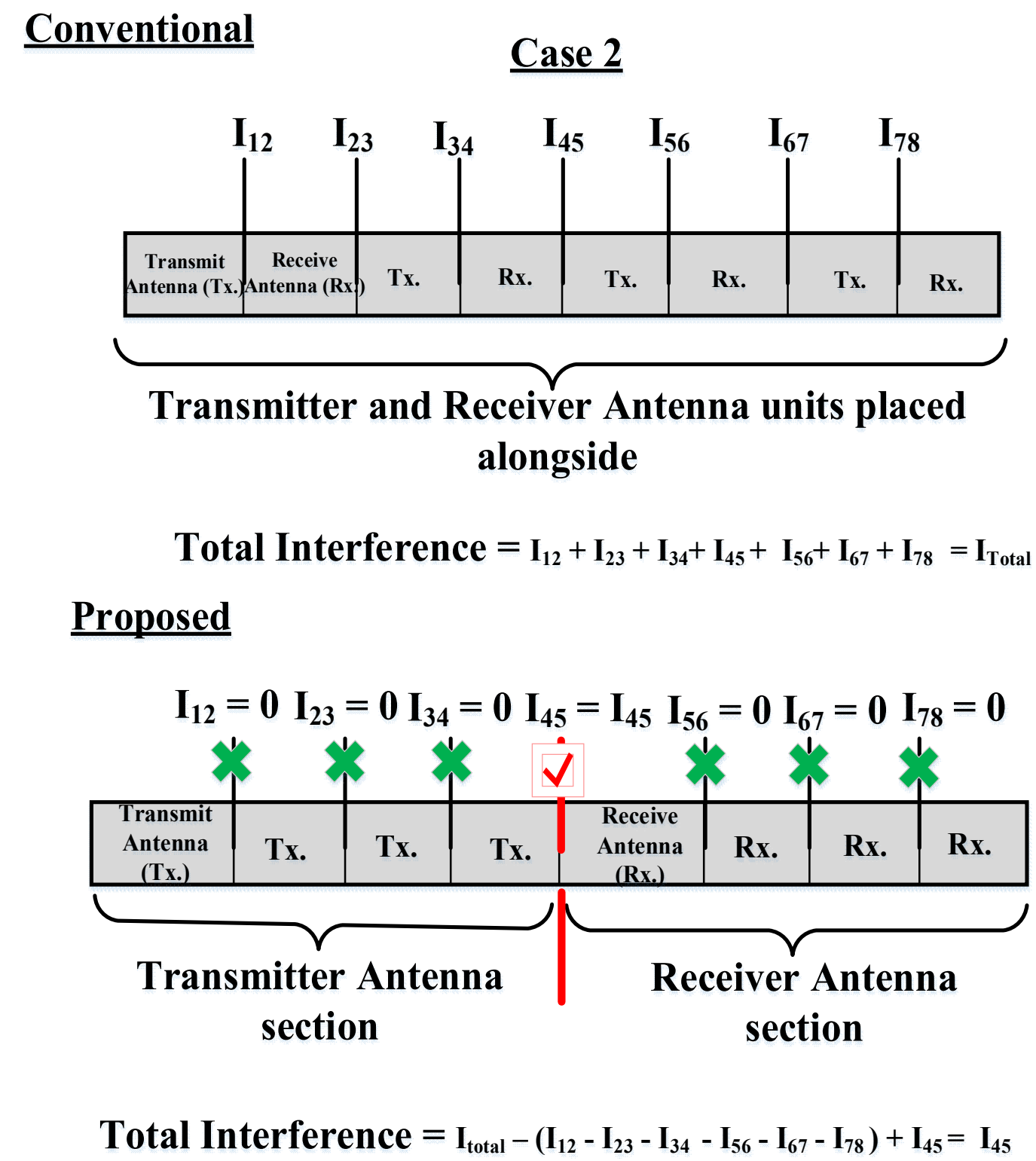

Fig. 8 Proposed case scenario 2 analysis.

### *C. Case 3: Separate Transmit and Receive Antenna Section with Offloading two Transmit Antennas*

This case is an amalgamation of both Case 1 and Case 2 scenarios. In this case we divide the transmitter and receive antenna section into two to mitigate interference as well as offload two transmit antennas.

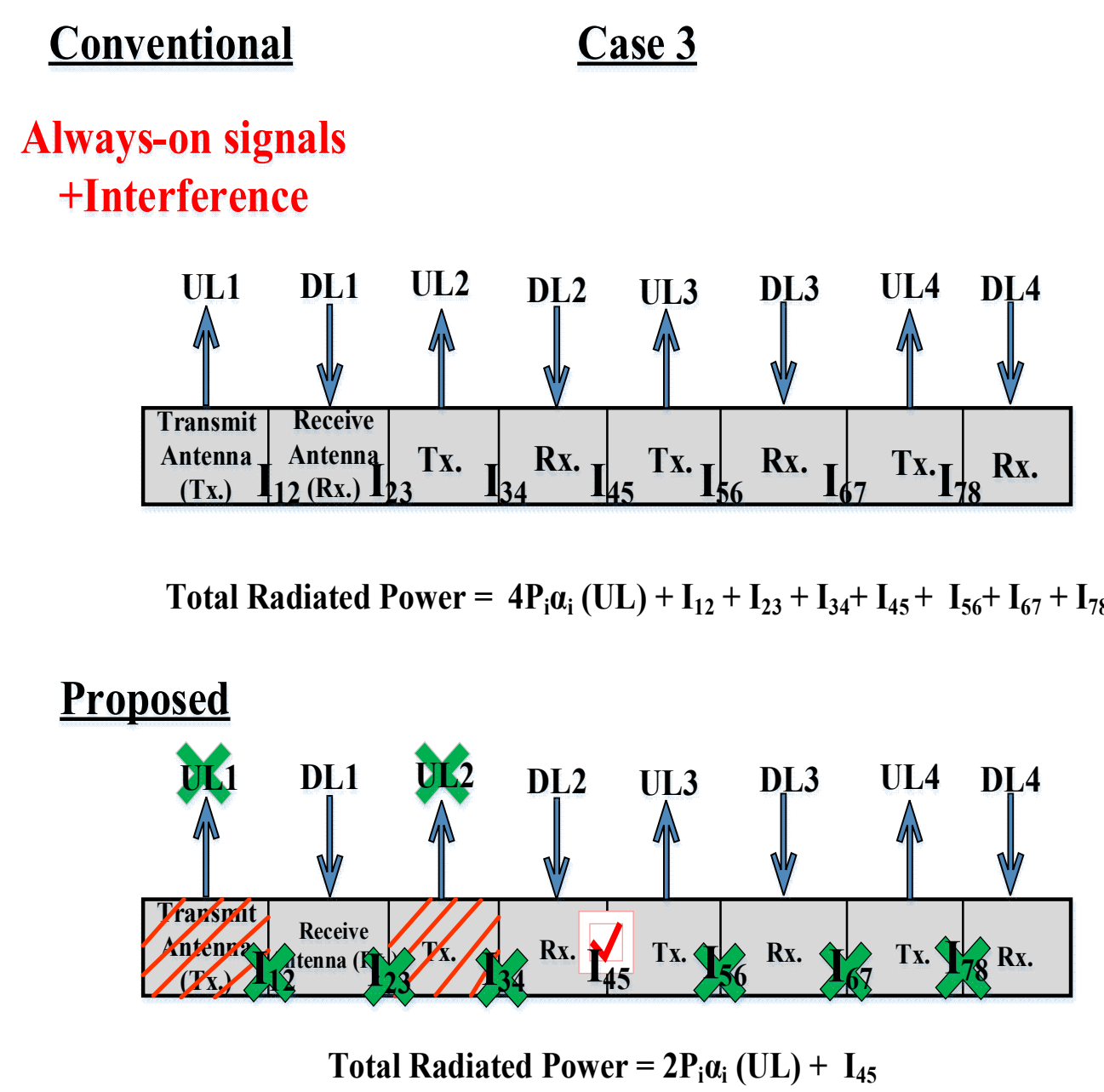

Fig. 9 Proposed case scenario 3 analysis.

This scenario provides the maximum output in case of power consumption and EM radiation reduction. The total power saved in this scenario is an additive of power conserved in offloading two uplink signals and canceling the individual interference between each transmitter and receiver units.
The total radiated power in both conventional and proposed case 3 is depicted in Fig. 9 and can be formulated as follows in equation (31) and (32) respectively:

$$P_{rad}^{c} = 4P_i\alpha_i(UL) + I_{12} + I_{23} + I_{34} + I_{45} + I_{56} + I_{67} + I_{78} \quad (31)$$

Radiated power in proposed case 3:

$$P_{rad}^{proposed} = 2P_i\alpha_i(UL) + I_{45} \quad (32)$$

## IV. EM RADIATION METRIC CALCULATION FOR PROPOSED METHODOLOGY

In the previous sections we have calculated the power radiated in case of each of the antenna topologies. This radiated power is incident on the body and causes local heating of the exposed body tissue. We study the impact of EM radiation on body tissues with the help of EM radiation metrics. There are various metrics such as Power Density (PD), Specific Absorption Rate (SAR), Exposure Ratio (ER), transient and steady state temperature elevation. Power density is calculated as a ratio of the radiated power by an antenna to the area covered by the radiation beam pattern of the radiating antenna.
The power density for the radiated power from the antennas can be computed from the following expression:

$$\text{Power Density } PD = \frac{G_{tr}\, P_{rad}}{4\pi d^2} \quad (33)$$

Power density for 4-antenna or 8-antenna system can be computed as additive sum of individual power densities for each of the radiating antenna.
The Power density radiated form each antenna for 4-antenna array system is formulated as follows:

$$PD_{4,ant\ array} = \frac{G_{tr}\,P_1}{4\pi d^2} + \frac{G_{tr}\,P_2}{4\pi d^2} + \frac{G_{tr}\,P_3}{4\pi d^2} + \frac{G_{tr}\,P_4}{4\pi d^2} \quad (34)$$

Similarly, the power density for 8-antenna system can be written as following:

$$PD_{8,ant\ array} = \frac{G_{tr}\,P_1}{4\pi d^2} + \frac{G_{tr}\,P_2}{4\pi d^2} \ldots\ldots\ldots \frac{G_{tr}\,P_8}{4\pi d^2} \quad (35)$$

Specific Absorption rate is calculated with respect to the radiation intensity that is incident directly on a body tissue. Higher the radiation exposure more is the value of SAR obtained. It is the ratio of the exposed power incident on a tissue (measured in Watt) to the mass of the tissue in Kilograms (Kg). If the number of active antennas in a mobile phone is more, the radiated power is high as the EM radiation exposure will rise. In our proposal we reduce the number of active antennas i.e. “always-on” signals, without compromising with the QoS and QoE( Quality of Experience). This helps in limiting the thermal heating of the exposed tissue decreasing the exposure ratio along with Specific Absorption Rate (SAR) in human body tissue.
SAR computation can be done as follows

$$SAR = \frac{P_{radiated}}{Mass\ of\ Tissue\ (Kg)} \quad (36)$$

SAR for 4-antenna system is formulated as follows:

$$SAR_{4,ant\ array} = \frac{P_1 + P_2 + P_3 + P_4}{Mass\ of\ Tissue\ (Kg)} \quad (37)$$

SAR for 8-antenna system is formulated as follows:

$$SAR_{8,ant\ array} = \frac{P_1 + P_2 + P_3 + P_4 \ldots\ldots P_8}{Mass\ of\ Tissue\ (Kg)} \quad (38)$$

Exposure Ratio (ER) is another metric which is a ratio of the electric field produced by a device to the reference level of electric field or Maximum permissible exposure allowed in that area. Mathematically it can be represented in the form of power density as follows:

$$ER = \vartheta\left(\frac{PD_{signal}}{PD_{total}}\right) \times 100\% \quad (39)$$

where $PD_{signal}$ is the power density of the RF signal and $PD_{total}$ is the cumulative power density of all the signals at a particular geographical location $\vartheta$.
The last EM metric discussed here is steady state and transient temperature elevation. It is the rise in the temperature of a body tissue which is exposed directly to radiated power from a device. The rise in temperature is calculated in degree Celsius with the help of Pennes bioheat equation. It incorporates various aspects to calculate the elevation in temperature such as tissue thickness, mass density, specific heat capacity.

## V. SIMULATIONS AND DISCUSSIONS

The simulations have been performed in MATLAB software to analyze the performance in terms of different parameters. We calculate the achieved capacity i.e. QoS (Quality of Service), directivity and biological impact of radiation caused with different antenna topologies. The analysis is performed for 5G NR frequency bands i.e. sub-6GHz and mmWave band keeping in view the channel consistency for 6G [21]. The simulation parameters used in this work have been obtained from the existing works in literature that study the EM radiation impact biologically. To study the performance of different antenna elements in terms of achieved capacity we calculate the ergodic capacity for varying antenna elements. The ergodic capacity is computed for 2 element array, 8 element array and our proposed 6 element array. This analysis has been performed over the frequency range of 1-6 GHz i.e., sub-6GHz band. We assume CSIT condition i.e., there is complete knowledge of channel state information at the transmitter. A comparative graph is shown in Fig. 10 depicting the achieved capacity over the range of frequency for the three antenna topologies.

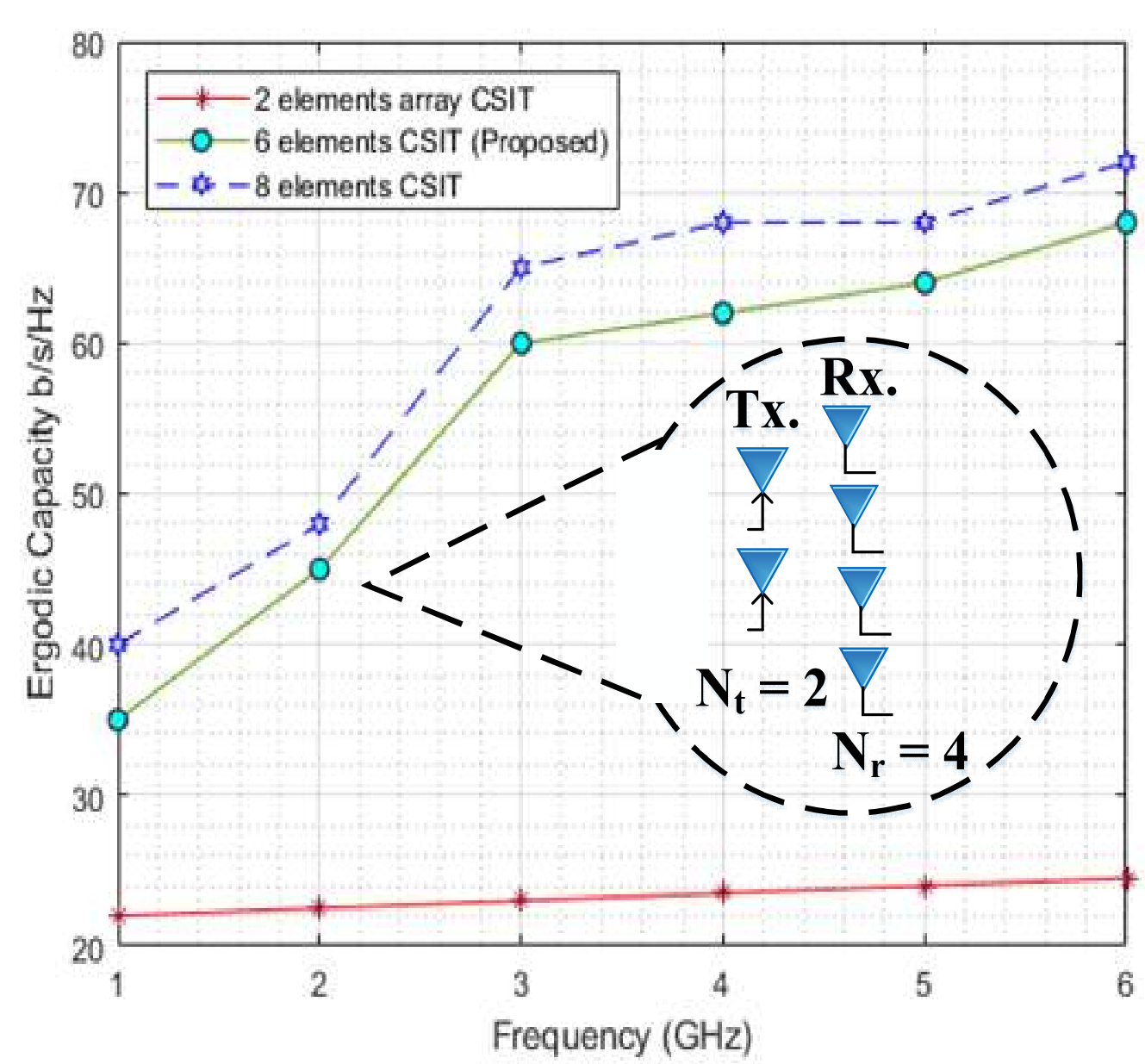

Fig. 10 Ergodic Capacity v/s Frequency.

From the graph we obtain that maximum ergodic capacity is achieved for 8 antenna array and least for 2 elements array. The proposed 6 element array with 2 transmit antennas and 4 receive antennas also achieves high capacity that achieves the target rate with power consumption as we offload 2 transmit antennas in proposed topology. The reduction in EM radiation of proposed antenna topology is validated with the value of power density obtained.

We calculate the value of power density w.r.t frequency as well as Skin depth as shown in Fig. 11. Both far-field and near-field values of power density are computed. In Fig. 11 a) the far-field value of power density v/s frequency is computed for 8 element array and proposed 6 element array. It is evident from the plot that the proposed antenna configuration produces less power density by offloading two transmit antennas.

In case of a device with all antennas active and communicating with Wi-Fi communication the power density obtained is higher and increases with rise in the frequency of operation. The power saved when two uplink information transfer signals are inactive helps in conserving power and hence the emitted radiation is reduced. This aids in overall reduction in the power density obtained as seen in mathematical modelling in previous section. In Fig. 11 b) the near-filed exposure of power density is calculated w.r.t skin depth. It is calculated for sub-6GHz frequency band. This graph depicts the comparison of 8 element array and proposed 6 element array for the three case scenarios. It is found that highest power density is obtained for 8 element array which decreases with increase in skin depth as maximum radiation is absorbed in the upper layers i.e., epidermis and dermis layers. For the proposed case scenarios there is significant reduction in power density radiation with least radiation for case 3.

Since case 3 includes both case 1 and 2 there is maximum power saved. For case 1 and 2 there is not much difference in power saved because the power saved in offloading two transmit antennas is almost equivalent to the reduction in interference power obtained in case 2 by separating transmit and receive antenna section. Another radiation metric used to study the EM radiation impact is Exposure Ratio (ER). It is used to study the impact of emitted radiation from a device for a particular duration and skin depth. We have obtained a 3D comparison of exposure ratio for 8 element arrays with all the three case scenarios. Fig. 12 depicts the graphical depiction of all the comparisons and case scenarios for values of ER.

The values of ER obtained are for a particular duration of time considering different antenna topologies.The first graph consists of 3D plot obtained for 8 element array and proposed 6 element array (Case 1). The value of ER obtained for 8 element array is about 0.8 and that for proposed case 1 is 0.4 at a skin depth of 10mm and duration of 40 sec. The computation of ER is carried out at sub 6-GHz frequency. With increase in skin depth the value of ER deceases for both the scenarios. For case 2 the ER value is further reduced to 0.31 which is the maximum value at a duration of 40 sec and 10mm skin depth. For case 3 the maximum value obtained is 0.13 as can be seen from the 3D plot obtained with maximum value at a duration of 40 sec. In the last graph we obtain a comparative graph of all the proposed case scenarios to obtain a visualization of all the ER values obtained. It can be seen that maximum value is obtained for case 1, then case 2 and lowest for case 3 according to increase in the power being saved respectively.

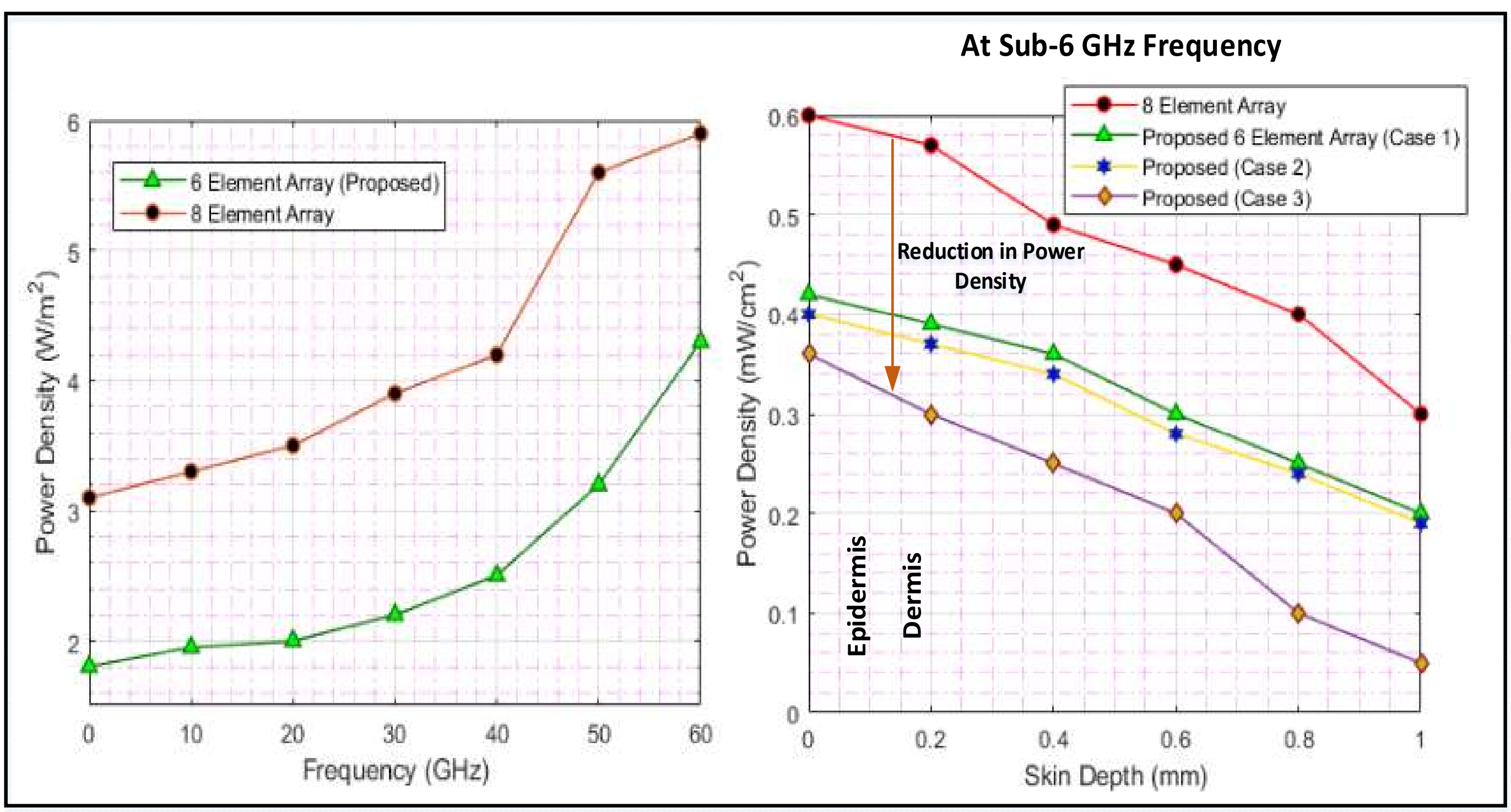

Fig. 11 a.) Power Density v/s Frequency (Far-Field exposure) b.) Power Density v/s Skin Depth. (Near-Field exposure)

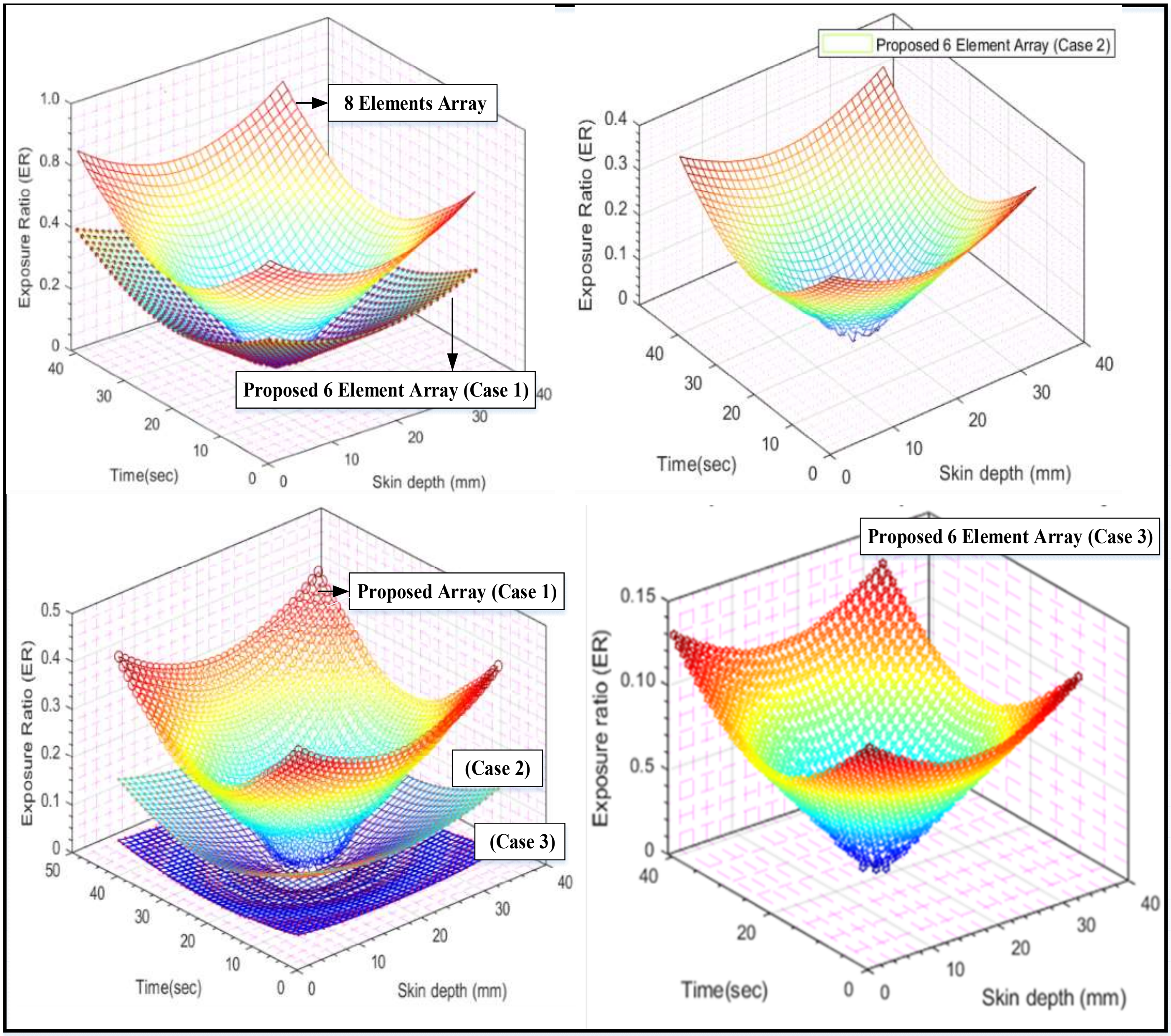

Fig.12 Exposure Ratio for 8 Element and proposed 6 Element array system at Sub-6 GHz frequency.

We also obtain the specific absorption rates for 8-element and proposed case scenarios w.r.t skin depth. SAR is a near field radiation metric that measures the radiation absorbed in the exposed body tissue due to local heating as a result of near-field exposure [22].

The graph obtained in Fig. 13 depicts SAR v/s skin depth for 8 element array and proposed 6 element array. In the first graph the data tips denote the values for SAR at a skin depth of 0.2 mm. It is evident that a lower value of SAR is obtained for 6 element array system for the same skin depth because the radiated power is less in the proposed case. For the second graph a comparison of 8 element array is made with all the case scenarios. The data tips are highlighted for each topology at a skin depth of 0.2mm to clearly demarcate the obtained values of SAR for each case. The power saved in each case reduces the obtained SAR which decreases with increase in skin depth.

A comparison table is given in Table I depicting the numerical values of the SAR obtained, comparing it with existing 8 Element array system and the 3 proposed case scenarios. We have also obtained a comparison for directivity at different frequencies i.e., 3.5 GHz, 28 GHz and 54 GHz. The directivity values are obtained for 2 element, 8 elements and proposed 6 element arrays as in Fig.14. It is found that with the increase in operating frequency and number of antenna elements the value of directivity also increases. At a particular frequency such as 3.5 GHz maximum directivity is obtained for 8 element array and same pattern is observed at other frequencies also. Higher the number of antenna elements more is the diversity obtained and hence spatial multiplexing. Antenna designs must be resistant to structural and lossy human body deformations.

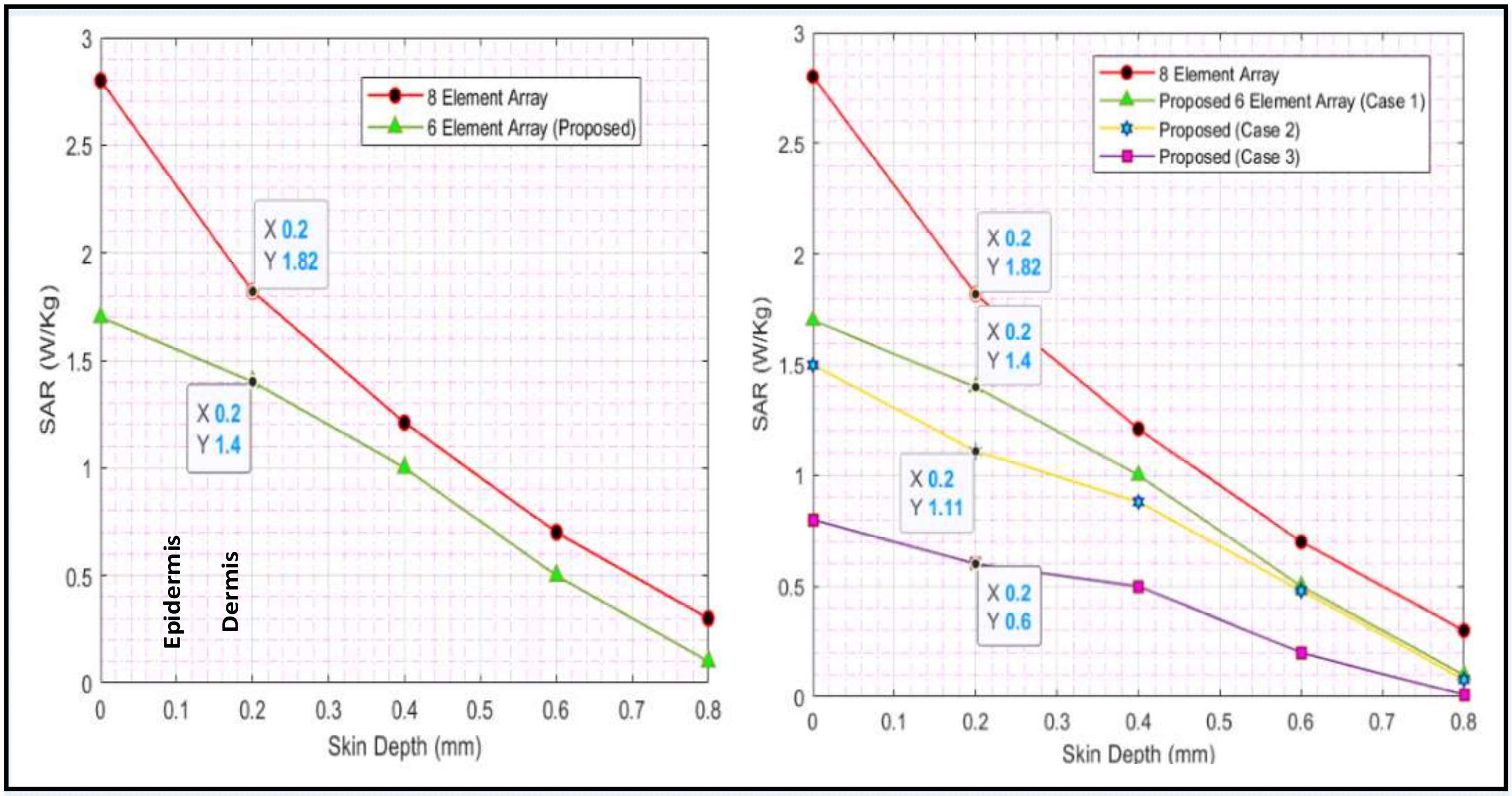

Fig.13 SAR v/s Skin Depth.

TABLE I
COMPARISON OF SAR VALUES FOR CURRENT ANTENNA TOPLOGY AND PROPOSED CASE SCENARIOS

| S.No. | Skin Depth (mm) | Specific Absorption Rate (SAR) (Watt/Kg) | | | |
|---|---|---|---|---|---|
| | | 8 Element Array (Currently existing) | Proposed (Case 1) | Proposed (Case 2) | Proposed (Case 3) |
| 1. | 0 | 2.75 | 1.71 | 1.5 | 0.8 |
| 2. | 0.2 | 1.82 | 1.4 | 1.11 | 0.6 |
| 3. | 0.5 | 1.25 | 1.10 | 0.90 | 0.5 |
| 4. | 0.6 | 0.75 | 0.70 | 0.69 | 0.22 |
| 5. | 0.8 | 0.31 | 0.22 | 0.20 | 0.09 |

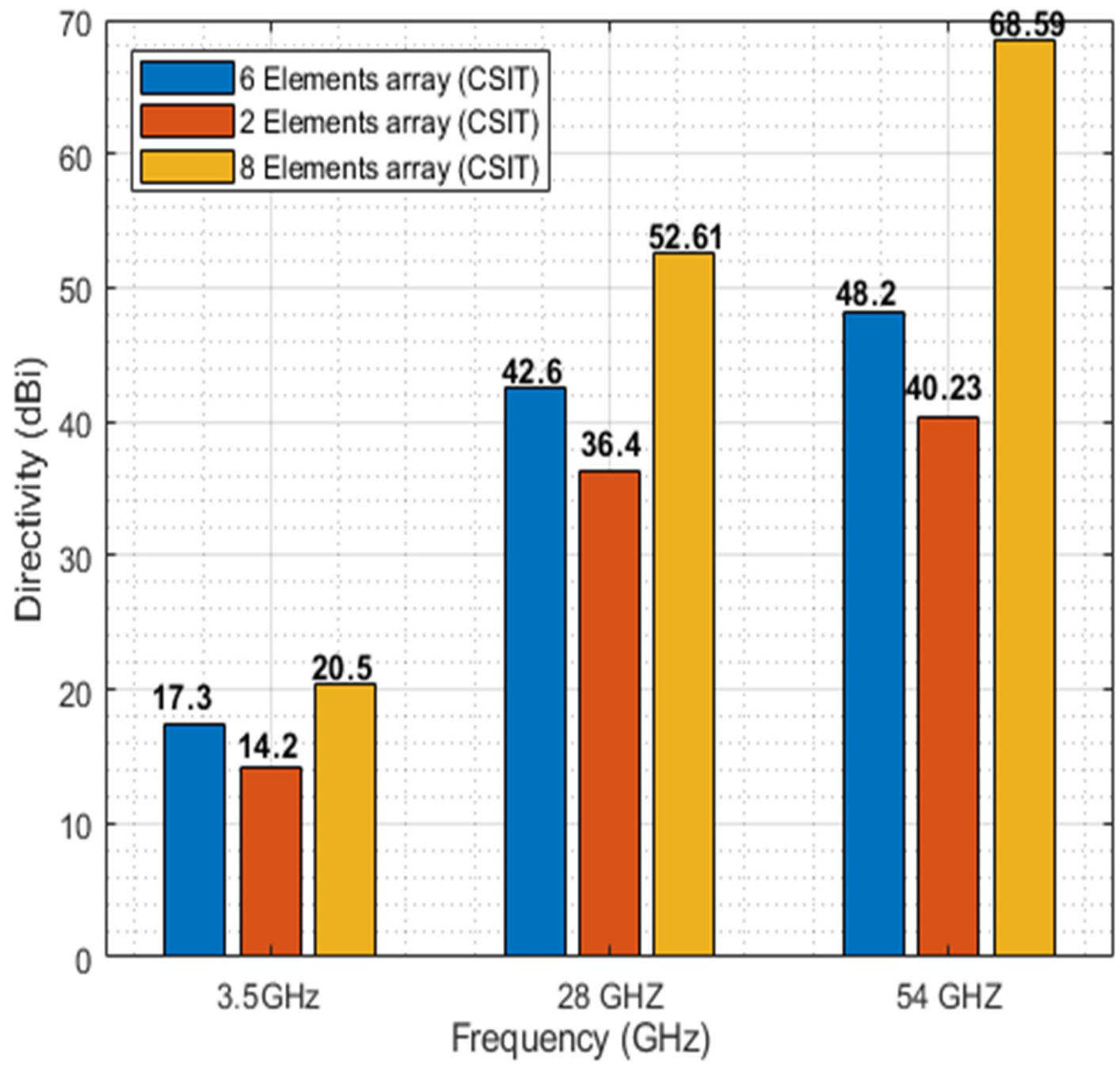

Fig. 14 Directivity v/s Frequency.

## VI. CONCLUSION

In this article we present an antenna topology for 6G smart phones for future cellular networks. In the proposed topology we have proposed three case scenarios with the basis of 8 element array. In the first case we offload two transmit antennas when the mobile user is operating over Wi-Fi to limit the always-on radiations/signals. In the second case we manage the interference by separating the transmitter and receiver antenna section. In the third case we incorporate both case 1 and 2 collectively. The three case scenarios have been proposed keeping in view the EM radiation impact produced by cellular devices with its always ON antennas. As we will be including mmWave frequencies for operation for future 6G mobile and wireless communication, it is imperative to study the aspect of biological safety.

We have studied EM radiation metrics for our proposal to validate its performance in radiation reduction. The comparison has been made using power density, SAR, exposure ratio. From the simulations and results obtained it has been verified that the proposed antenna configuration is much safer in operation and also achieved desired data rate without compromising on the QoS and QoE. In terms of near field radiation metric the percentage reduction in SAR achieved is about 71% in comparison to existing scenario and 84% reduction in exposure ratio has been achieved with proposed antenna topologies. The power density reduction achieved is about 40% for current topology and proposed antenna topology (case 3). The antennas topologies that radiate the least amount of Electromagnetic radiation whilst achieving the target user requirements is the need of the hour for future 6G hand held devices.

## ACKNOWLEDGMENT

The authors gratefully acknowledge the support provided by Satish Dhawan Centre for Space Sciences, ISRO, Central University of Jammu, Jammu and Kashmir, India. and Communication Lab at IIIT Jabalpur, India.

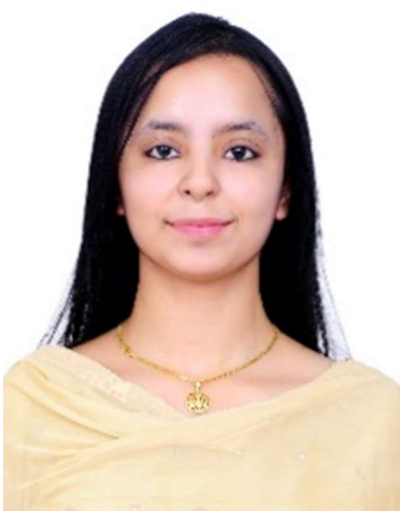

HANEET KOUR is currently working as Research Associate at SDCSS, ISRO, Central University of Jammu, J & K. She has completed her Ph. D degree in Electronics and Communication Engineering from Shri Mata Vaishno Devi University, Katra, Jammu and Kashmir, India. Her research interest includes the emerging technologies of 5G/6G wireless communication networks, power optimization, Green Communication, Network Safety/ Reliability and Satellite Communication. She is working on MATLAB tools for Wireless Communication, has received student travel grant from COMSNET in 2019 and 2020. She is a student member of Institute of Electrical and Electronics Engineers (IEEE).

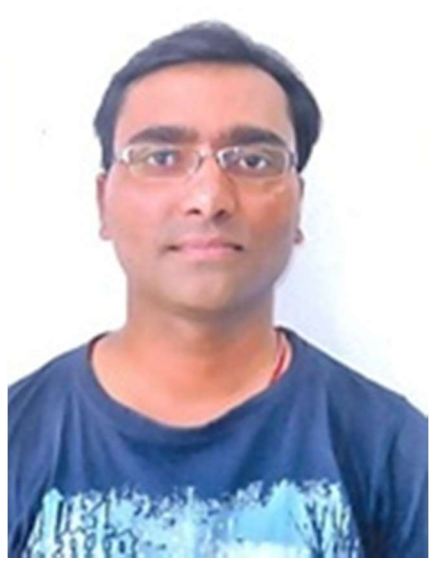

Dr. RAKESH K JHA (S'10, M'13, SM 2015) is currently an Associate Professor in the Department of Electronics and Communication Engineering, Indian Institute of Information Technology Design and Manufacturing Jabalpur, India. He has also worked as an Associate Professor at SMVD University, J&K, India. He is among the top 2% researchers of the world. He has published more than 61 SCI Journals Papers including many IEEE Transactions, IEEE Journal, and more than 25 International Conference papers. His area of interest is Wireless communication, Optical Fiber Communication, Computer Networks, and Security issues. Dr. Jha's one concept related to the router of Wireless Communication was accepted by ITU in 2010. He has received the young scientist author award by ITU in Dec 2010. He has received APAN fellowship in 2011, 2012, 2017 and 2018 and student travel grant from COMSNET 2012. He is a senior member of IEEE, GISFI and SIAM, International Association of Engineers (IAENG) and ACCS (Advance Computing and Communication Society). He is also a member of, ACM and CSI, with many patents and more than 6001 citations to his credit.

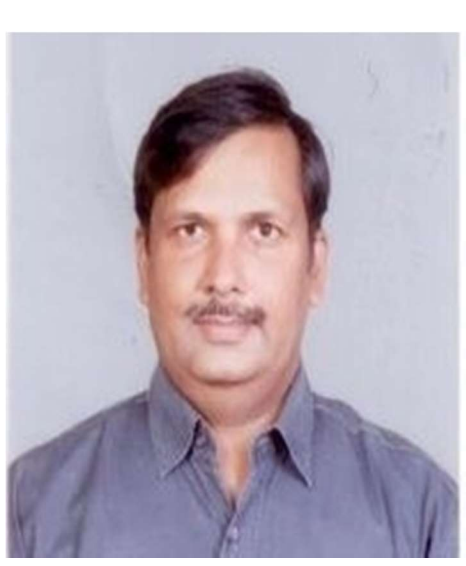

PROF. SANJEEV JAIN, Prof. Sanjeev Jain, born at Vidisha in Madhya Pradesh in 1967, obtained his Post Graduate Degree in Computer Science and Engineering from the Indian Institute of Technology, Delhi, in 1992. He later received his Doctorate Degree in Computer Science & Engineering and has over 24 years' experience in teaching and research. He has served as Director, Madhav Institute of Technology and Science (MITS), Gwalior. and as Director IIITDM, Jabalpur. He has also served at SMVDU, Katra as vice-chancellor. Presently he is serving as Vice- Chancellor at Central University of Jammu, J & K. Besides teaching at Post Graduate, level Professor Jain has the credit of making significant contributions to R & D in the area of Image Processing and Mobile Adhoc Network.